\documentclass[%
 reprint,
superscriptaddress,
showpacs,preprintnumbers,
nofootinbib,
 amsmath,amssymb, 
 aps,
 prd,
 longbibliography,
]{revtex4-1}

\usepackage{booktabs}
\usepackage{graphicx}
\usepackage{mathrsfs}
\usepackage{enumitem}
\usepackage{comment}
\usepackage{xcolor}

\graphicspath{{figs/}}

\begin{document}

\title{A composite solution to the EDGES anomaly}
\author{Anubhav Mathur} 
\affiliation{Department of Physics and Astronomy, Johns Hopkins University, 3400 N. Charles St., Baltimore, MD 21218, USA}
\author{Surjeet Rajendran}
\affiliation{Department of Physics and Astronomy, Johns Hopkins University, 3400 N. Charles St., Baltimore, MD 21218, USA}
\author{Harikrishnan Ramani}
\affiliation{Stanford Institute for Theoretical Physics, Stanford University, Stanford, CA 94305, USA}
\date{September 2021}

\begin{abstract}
Subcomponent millicharged dark matter that cools baryons via Coulomb interactions has been invoked to explain the EDGES anomaly. However, this model is in severe tension with constraints from cosmology and stellar emissions. In this work, we consider the consequences of these millicharged particles existing in composite Q-balls. The relevant degrees of freedom at high temperature are minuscule elementary charges, which fuse at low temperatures to make up Q-balls of larger charge. These Q-balls serve as the degrees of freedom relevant in cooling the baryons sufficiently to account for the EDGES anomaly. In such a model, cosmology and stellar constraints (which involve high-temperature processes) apply only to the feebly-interacting elementary charges and not to the Q-balls. This salvages a large range of parameter space for millicharged Q-balls that can explain the EDGES anomaly. It also opens up new parameter space for direct detection, albeit at low momentum transfers. 
\end{abstract}
\maketitle 
\section{Introduction}

The Experiment to Detect the Global Epoch of Re-ionization Signature (EDGES) experiment has reported a dip in the 21 cm spectrum corresponding to strong absorption around $z=17$~\cite{Bowman:2018yin}. This can be interpreted as a 3.8 $\sigma$ deviation from the  $\Lambda$CDM prediction for the baryon temperature~\cite{Barkana:2018lgd, Barkana:2018qrx, Kovetz:2018zan}. Dark matter (DM) cooling of the baryonic fluid has been invoked as an explanation for this excess~\cite{Barkana:2018lgd, Barkana:2018qrx,Kovetz:2018zan}. A DM model that maximizes the cross-section around cosmic dawn is sub-component millicharged dark matter (mCDM), which has a larger cross-section with Standard Model (SM) charges at the lowest relative velocities. However, the millicharge parameter space is extremely constrained due to limits from CMB and BBN, cooling of SN1987A and stars, and terrestrial experiments~\cite{Kovetz:2018zan}. It has subsequently been shown that even this limited parameter space results in over-production of mCDM through freeze-in~\cite{Berlin:2018sjs, Creque-Sarbinowski:2019mcm}. 

These difficulties have led to two other ways to solve the EDGES anomaly. The first involves heating the CMB relative to baryons \cite{Feng:2018rje,Ewall-Wice:2018bzf,Pospelov:2018kdh, Fialkov:2019vnb}, while the second involves mCDM which is tightly coupled to an additional cold component that forms the dominant DM which does the bulk of the cooling~\cite{Liu:2019knx}. In this paper, we point to a third possibility. The mCDM explanations for the EDGES anomaly to date have treated the millicharged particle (mCP) as elementary without internal structure. As a result, the same mCP is the physical particle at all energies. In this work, we explore the consequence of this mCP being a composite state of elementary mCPs with much smaller mass and charge, glued together by a force that confines at low temperatures. The elementary charges are the relevant degrees of freedom at temperatures and energies much higher than cosmic dawn. As a result, in our model, constraints from CMB, BBN, over-closure, stellar and SN cooling as well as colliders all apply only to the elementary charges. We demonstrate here that there is a drastic increase in new parameter space for mCDM as long as it is in a composite state. Furthermore, we explore the unique thermal history for the dark sector that involves confinement when $T_D$ (the dark temperature) falls below $\Lambda_D$ (the dark confining scale) and deconfinement if the dark temperature increases subsequently. We point out a novel dark phase where thermal contact with the SM results in a thermostatic dark bath, i.e. the dark bath staying at the same temperature with the heat dump from baryons exactly cancelled by Hubble cooling.
 
 \section{Model}
We consider mCP scalars $s$ that carry electric charge $\epsilon_s$ and mass $m_s$\footnote{As we explain later, fermionic mCPs share many of the properties of scalar composites discussed in this work via nucleosynthesis. However, the Fermi temperature of the constituents exceeds the cold temperatures required to explain the EDGES anomaly.}.
  The scalar is assumed to have quartic and effective sextic self-interactions. These self-interactions make it possible for the scalars to form Q-balls {\it i.e. } these Q-balls will be the composite states of interest to us.  We also assume that the mCPs are charged under a dark $U(1)$ with a dark charge $g_D$. The interactions of the scalar field $s$ are captured in the Lagrangian
\begin{equation}
\mathcal{L}=\frac{m_s^2 |s|^2}{2} -\lambda |s|^4 + \frac{1}{\Lambda_6} |s|^6 +D_\mu s D^\mu s^* +\epsilon F^{\mu \nu} F'_{\mu \nu}
\end{equation}
Here $D_\mu=\partial_\mu - i g_D A'_\mu$, and $F$ and $F'$ are the field strength tensors of the SM U(1) and the dark U(1) i.e. $A'$ respectively. This dark $U(1)$ accomplishes two goals: first, it allows the electric charge of the mCP to be generated via kinetic mixing and second, it provides Coulomb repulsion that prevents the Q-balls from getting too big. The mass of the dark photon is unimportant as long as it is sufficiently long ranged to allow the Q-balls to interact with baryons via Coulomb scattering. We can take this mass to be zero, or low enough to evade direct stellar constraints on the dark photon.  

For simplicity we will assume that these elementary mCPs that confine all have the same charge sign. For net charge neutrality, we envision an asymmetric dark component with the opposite charge, just like the standard model. Under the correct circumstances, the scalars fuse to form composite Q-balls or Q-balls $\chi$ with ``atomic number" Q such that they carry charge $\epsilon_\chi = Q \epsilon_s$. The mass of these Q-balls is given by 
\begin{equation}
m_\chi= Q\sqrt{ m_s^2-\frac{\lambda^2 \Lambda_6^2}{2}} = Q \Lambda_D
\end{equation}
i.e. the difference between the total mass square of the constituents minus the square of the binding energy. We assume $m_s \gtrsim \frac{\lambda \Lambda_6}{2}\approx\Lambda_D$ and $\Lambda_D$ can be viewed as the binding energy per scalar parton. 

The volume of these Q-balls is given by
\begin{equation}
V_\chi= \frac{Q}{\sqrt{m_s^2-\frac{\lambda^2 \Lambda_6^2}{2}}} \frac{2}{\lambda \Lambda_6^2}\approx \lambda \frac{Q}{\Lambda_D^3}
\label{volume}
\end{equation}
Thus the volume of these Q-balls can be parametrically smaller than their fermionic cousins, the nuclei $N$ which scale as $V_N= \frac{Q}{\Lambda_D^3}$. 
Finally, the Q-balls are stable only for large enough $Q$, where the surface tension term can be ignored, this gives \cite{Grabowska:2018lnd}

\begin{equation}
    Q_{\rm min}\approx \frac{1}{\lambda}
    \label{Qmin}
\end{equation}

\subsection{Synopsis of Q-ball Evolution } 
\label{sec:synopsis}
We start tracking the dark bath at temperature $T_D \gtrsim \Lambda_D$ at $z=1000$, with Q-balls broken into their constituent particles. At early times, the SM baryon temperature $T_b \gg \Lambda_D$. The constituents have low enough charge that they are not in thermal equilibrium. At late times, they undergo a phase transition and rapidly form small Q-balls which are assumed not to be in thermal equilibrium with the standard model \cite{Kusenko}. These small Q-balls will begin to fuse and form larger Q-balls provided the inter-particle spacing between Q-balls is larger than the size of Q-balls i.e. $n_{\rm Q-ball}^{-1}>V_{\rm Q-ball}$. This is satisfied as long as the phase transition happens later than
\begin{align}
    z_{\rm tr,max} &= \left(\frac{m_{\chi}}{f_{D}\rho_{\text{DM},0}}\frac{\Lambda_{D}^{3}}{\lambda Q}\right)^{1/3} \nonumber \\
    &\approx 24 \left(\frac{\Lambda_{D}}{1\text{ K}}\right)^{4/3} \left(\frac{f_{D}}{0.4\%}\right)^{-1/3} \left(\frac{\lambda}{10^{-7}}\right)^{-1/3}
    \label{zmax-tr}
\end{align}
It is important to note that the analogous calculation for fermions rather than Q-balls (corresponding to $\lambda=1$, as discussed above) forces the bound states to form much later than cosmic dawn, ruling out the entire parameter space. Stated differently, depending on their temperature in the early universe, throughout their evolution fermionic mCPs are either cold enough that they constitute a degenerate gas (in which case the composite states considered in this work are not the relevant degrees of freedom), or so warm that they cannot cool the baryons sufficiently to explain EDGES.

We choose the initial dark temperature of the Q-balls at $z=1000$ to be high enough that the phase transition happens after $z_{\rm tr,max}$. The requirement that the first Q-balls formed are (i) small enough to be out of thermal equilibrium with SM baryons at this redshift and (ii) stable at their size sets
\begin{equation}
    \lambda > 7\times10^{-8} \left(\frac{m_{\chi}}{\text{MeV}}\right)^{-1/5} \left(\frac{\Lambda_{D}}{1\text{ K}}\right)^{1/5} \left(\frac{f_{D}}{0.4\%}\right)^{-1/5}
    \label{lambda-min}
\end{equation}

As the Q-balls become bigger, they will interact with the SM baryons and start extracting energy from the standard model, leading to heating of the Q-ball sector. But the temperature of this sector cannot exceed $\Lambda_D$ since this results in Q-ball fission, leading to loss of thermal contact with the baryon bath, which leads to Hubble cooling. Once $T_b$ drops sufficiently that it is unable to transfer enough heat to hinder Q-ball growth, there is rapid fusion resulting in large Q-balls. The maximum size of the Q-ball in this case is set by Coulomb repulsion---as the Q-ball becomes larger, the repulsion from the dark $U(1)$ grows and it inhibits Q-ball growth beyond a certain size. Parameters are chosen so that this phase of Q-ball formation occurs around the redshifts of interest to the EDGES experiment. At this stage, the Q-balls scatter with the SM baryons, cooling the SM baryons and explaining the EDGES observations. 

\subsection{The Size of the Q-ball}
\label{sec:analysis}

It is necessary to limit the size of the Q-balls so that they can coherently scatter with the baryons as well as provide enough heat capacity in the Q-ball sector to cool the SM bath. The dark $U(1)$ provides the Coulomb repulsion necessary to enforce this limit. Since $g_D \gg \epsilon_s$, we will ignore the Coulomb repulsion from electromagnetism in this section. 

How do the Q-balls form? We follow the prescription developed in ~\cite{Krnjaic:2014xza}, to account for the Coulomb repulsion due to the dark $U(1)$. The ratio of fusion rate to Hubble rate for $\{Q_1,Q_2\}$ fusion is

\begin{equation}
\frac{n \sigma v}{H}=\frac{10^{32} f_D \lambda^\frac{2}{3} z^\frac{3}{2}}{Q^{\frac{5}{6}}}\left(\frac{10~{\rm K}}{\Lambda_D} \right)^2 \sqrt{\frac{T_D}{10~{\rm K}} \left(\frac{10~{\rm K}}{m_s}\right)^3}\gg 1 \nonumber
\label{Q-ballform} 
\end{equation}

As the Q-balls grow bigger, there is an increased Coulomb barrier to fusion as treated in ~\cite{Krnjaic:2014xza}. As the Q-balls become larger, this cross-section is suppressed by the Coulomb barrier. This is captured by $P_G(T)$ the Gamow factor~\cite{gamow1963quantum} which is the temperature dependent factor that captures the effects of the Coulomb barrier. This factor is: 
\begin{align}
    P_G(T)=e^{-G_E}=e^{-\sqrt{\frac{2\mu Q_1^2 Q_2^2 g_D^4}{T_D}}}
\end{align}
Here $Q_1$ and $Q_2$ are the sizes of the two Q-balls respectively and $\mu$ is their reduced mass. Thus fusion freeze-out depends critically on the Gamow factor. From the Gamow factor, it is clear that large-Q-ball small-Q-ball fusion will dominate over large-large type fusion due to weaker Coulomb repulsion. Moreover, as the Q-balls grow in size, the number density of larger Q-balls is lower than that of smaller Q-balls. Further the cross-section for a smaller Q-ball to merge with a larger Q-ball is set by the geometric size of the larger Q-ball. All of these factors imply that the growth of the Q-balls in our case is dominated by the mergers of small Q-balls with larger Q-balls. 

Let us now see how big these Q-balls can get {\it i.e.} estimate the freeze out of the fusion process. As seen in Eqn.~\ref{Q-ballform}, in the absence of the exponentially suppressed Gamow factor, the rate of the fusion process is very rapid compared to Hubble. The size of the Q-ball is then restricted purely by the exponential suppression from the Gamow factor which forces the process to freeze out. 

Taking $T_D\approx \mu \approx \Lambda_D$, for small-large fusion, the Gamow exponent is
\begin{equation}
 G_E\approx Q g_D^2 
\end{equation}
This places a bound on fusion growth,
\begin{equation}
    Q^{\rm lim}_{\rm Gamow}\approx g_D^{-2} 
\end{equation}
 This limit on the Q-ball size arising from the inhibition of their growth is the same ball-park as the stability limit~\cite{Grabowska:2018lnd} $ Q^{{\rm lim}}_{\rm stab}=1/g_{D}^{2}$ that can be placed on their size due to Coulomb repulsion. Q-ball freeze out occurs only due to the exponential dependence on Q-ball size in the Gamow factor. Thus, Q-balls whose sizes  are close to, but smaller than the Gamow limit are rapidly formed. This implies that as the universe expands and the temperatures drop, Q-balls will continue to grow until the Gamow limit is reached. 
 
It is also important to consider the heat that is released by the fusion process as the Q-balls grow.  Each fusion process releases roughly $\sim \Lambda_D$ per unit $Q$. Thus in the roughly $\sim Q$ fusion processes that occur to form a Q-ball of size $Q$, approximately $Q \Lambda_D$ energy into $\sim Q$ particles is released. Thus the heat released in the fusion cannot change the temperature by more than $\Lambda_D$ and thus does not hinder fusion.

While we expect this  mechanism to produce a range of Q-ball masses, the charge to mass ratio of all these Q-balls is the same. As seen later in Sec.~\ref{results}, for most of the relevant parameter space, results depend only on the charge to mass ratio, so it is justified to make a simplifying assumption that all Q-balls are of the same mass $m_\chi$. Note that this analysis of the Q-ball size is independent of the baryon temperature $T_b$. As we show in the following section,  $T_b$ is an important parameter in determining the number of mCPs that are fused into Q-balls but it does not determine the maximum size of a Q-ball. 

\subsection{Heat Transfer}
\label{sec:heat}
In order to understand heat transfer with the SM bath, we start by deriving the transfer cross-section for Q-balls to scatter with baryons. The differential cross-section for a mCP with charge $\epsilon$ to scatter with protons/electrons is \cite{Dvorkin:2013cea},
\begin{equation}
    \frac{d\sigma}{d\cos\theta}= \frac{2\pi\epsilon^2\alpha^2}{\mu^2 v_{\rm rel}^4(1-\cos\theta)^2}.
    \label{diffcross}
\end{equation}
with $\mu$ the reduced mass and $v_{\rm rel}$ the relative velocity. 

The forward divergence is cut off by the Debye mass of the mediator. For the SM photon, the Debye mass squared is given by
\begin{equation}
\Pi_A= e^2\left(\frac{x_e n_b}{T_{\rm b}} \right)
\end{equation}
where $x_e\equiv n_e/n_{\rm H}$ is the free-electron fraction, determined using~\cite{Hyrec1,Hyrec2}. The Debye mass is approximately $10^{-6}$ eV at $z=1000$ and $3\times 10^{-8}$ eV at $z=10$. The Debye mass square of the dark photon is 
\begin{equation}
\Pi_{A'}= g_D^2\left(\frac{n_D}{T_{\rm D}} \right)
\end{equation}
For the parameter space we are interested in, $g_D^2 n_D \ll e^2 x_E n_b$ and $T_D \le T_b$, such that $\Pi_{A'} \ll \Pi_{A}$. Hence we take only the SM photon Debye mass to regulate the divergence. Finally, for elementary charges, $q_{\rm max}=2\mu v_{\rm rel}$, such that the $\theta$ integral is taken between the limits $\theta= \{-1,\frac{2\epsilon \alpha \sqrt{\Pi_A}}{3T_b}\}$. For Q-balls, $q_{\rm max}\sim \text{Min}\left(R_{\rm Q-ball}^{-1},2\mu v_{\rm rel}\right)$, such that $\theta_{\min}=1-\frac{q_{\rm max}^2}{2\mu^2 v_{\rm rel}^2}$. 
The thermal-averaged transfer cross-section in the $q^2_{\rm max} \gg \Pi_A$ limit is given by integrating Eqn.~\ref{diffcross} over $\theta$, giving,

\begin{equation}
    \sigma_T = \frac{2 \pi \epsilon^2\alpha^2\xi}{\mu^2 v_{\rm rel}^4}
    \label{transfer}
\end{equation}
with $\xi=\ln\left({\frac{9 T_b^3}{4\pi \epsilon^2 \alpha^3 x_e n_b}} \frac{q_{\rm max}^2}{2\mu^2 v_{\rm rel}^2}\right)$. 
In the region of interest, it is safe to ignore the factor $\frac{q_{\rm max}^2}{2\mu^2 v_{\rm rel}^2}$ since it is inside the log.

Next, we compare the rate for Q-balls scattering off baryons to the Hubble rate:
\begin{equation}
    \frac{n_{b}\sigma_{T}v_{\text{rel}}}{H} \sim 10^{-18}\left(\frac{m_{\chi}}{\text{MeV}}\right)^{-1/2} \left(\frac{T_{D}}{10\text{ K}}\right)^{-3/2} \left(\frac{\epsilon_{\chi}}{10^{-14}}\right)^{2} z^{3/2}
\end{equation}

As a result, the smallest Q-balls with charges $\sim \epsilon_s$ which we take to obey stellar-cooling constraints discussed next in Eqn.~\ref{stellar}, are never in thermal contact with the SM. We also see that larger Q-balls with charge $\epsilon_\chi \gtrsim 10^{-7}$ can interact with the SM bath. At temperatures around $T_D\approx \Lambda_D$, both small Q-balls and large Q-balls can co-exist. Defining $\mathcal{F}_{\rm Q-ball}(z)$ as the mass fraction of large Q-balls i.e. Q-balls with $Q\sim Q^{\rm lim}_{\rm Gamow}$, 
\begin{align}
\dot{T}_D^{\rm ref}(\mathcal{F}_{\rm Q-ball}) =&-2 H T_D +\frac{2}{3}\frac{m_{\chi} x_e \rho_b }{\left(m_{\chi}+m_b\right)^2}\mathcal{F}_{\rm Q-ball} \frac{\sigma_0}{u_{\chi,b}^3} \nonumber \\ &\times \left\{ \sqrt{\frac{2}{\pi}} \left(T_b-T_D \right)\right\}
\label{thermeqn}
\end{align}

Here $u_{\chi,b}=\sqrt{\frac{T_D}{m_\chi}+\frac{T_b}{m_b}}$ is the average relative velocity due to thermal motion and $\sigma_0 = \sigma_T v_{\rm rel}^4$. We have verified that the bulk relative velocity between the $\chi$ bath and SM fluids does not contribute substantially to the thermal evolution of either fluid. When $T_D \gtrsim \Lambda_D$, the relevant degrees of freedom are the elementary charges, which have no thermal contact with the SM such that the dark fluid cools due to Hubble expansion. When $T_D$ drops below $\Lambda_D$ there is rapid Q-ball formation. 
These Q-balls can now interact with the SM and heat up, but the temperature cannot exceed $\Lambda_D$; after all, thermal contact with the SM would immediately be lost.
Consequently if the second term in Eqn.~\ref{thermeqn} dominates for $\mathcal{F}_{\rm Q-ball}\rightarrow 1$, then $\mathcal{F}_{\rm Q-ball}$ adjusts to smaller values so as to keep $\dot{T}_D=0$.  Thus, we set [Note: eqn below changed since initial dark temp is greater than Lambda, not zero.]

\begin{equation}
\dot{T}_D = 
\begin{cases}
\dot{T}_D^{\rm ref}(\mathcal{F}_{\rm Q-ball}=0) \quad & T_D > \Lambda_D  \\
\max\left(0, \dot{T}_D^{\rm ref}(\mathcal{F}_{\rm Q-ball}=0)\right) & T_D\leq\Lambda_D 
\end{cases}
\end{equation}

In the regime where $\dot{T}_D=0$, we can solve for the $z$ dependent fraction in Q-balls $\mathcal{F}_{\rm Q-ball}$ by setting Eqn.~\ref{thermeqn} to 0. We find for $T_D \leq \Lambda_D$,

\begin{align}
\mathcal{F}_{\rm Q-ball}  =&\textrm{Min} \left(1,2 H T_D \times \left[ \frac{2}{3}\frac{m_{\chi} x_e \rho_b }{\left(m_{\chi}+m_b\right)^2} \frac{\sigma_0}{u_{\chi,b}^3}\right. \right. \nonumber \\ &\left. \left. \times \left\{ \sqrt{\frac{2}{\pi}} \left(T_b-T_D \right)\right\}\right]^{-1}\right)
\label{fsolve}
\end{align}

We can see that even after the elementary charges have cooled to $\Lambda_D$ and formed Q-balls, in the limit where $T_b \gg T_D$ and when interactions are strong enough, the quantity in square brackets is much larger than Hubble cooling and hence $\mathcal{F}_{\rm Q-ball} \rightarrow 0$. This happens because in this limit, large Q-balls that form immediately break up into smaller ones that are not in thermal contact with the SM. As the disparity between $T_b$ and $T_D$ shrinks, $\mathcal{F}_{\rm Q-ball} \rightarrow 1$. 

The time evolution of the baryon temperature obeys
\begin{align}
\dot{T}_{\rm b} =&-2 H T_b +\frac{2}{3}\frac{m_{b} x_e \rho_D }{\left(m_{\chi}+m_b\right)^2}\frac{\mathcal{F}_{\rm Q-ball} f_D}{1+f_{\rm He}+x_e} \frac{\sigma_0}{u_{\chi,b}^3} \nonumber \\ &\times \left\{ \sqrt{\frac{2}{\pi}} \left(T_D-T_b \right)\right\}+\Gamma_C (T_{\rm CMB}-T_b)
\label{thermbary}
\end{align}
where $f_{\rm He}\equiv n_{\rm He}/n_{\rm H}$ is the helium fraction and $\Gamma_C$ is the Compton scattering rate.

While the initial temperatures of the CMB and baryon gas are set by observation, the initial dark temperature has to be chosen such that the Q-ball phase transition occurs no earlier than $z_{\rm tr,max}$. As described in Section~\ref{sec:analysis}, we find this value by incorporating Eqn.~\ref{lambda-min} into Eqn.~\ref{zmax-tr}, to ensure that the Q-balls are stable and out of thermal equilibrium with the baryonic gas when they first form. The initial dark temperature must also exceed $2\Lambda_{D}$ so that we always start off with Q-balls completely broken apart into their constituents. Combining these requirements yields the initial conditions
\begin{align}
T_b(z=1000)&=T_{\rm CMB} (z=1000)\approx T_{\rm CMB}^0 \times 1000 \nonumber \\
T_{\rm CMB}^0&= 2.725 ~\textrm{K}  \nonumber \\
T_D (z=1000) &= \max\left(2\Lambda_{D}, \left(\frac{1000}{z_{\rm tr,max}}\right)^{2}\Lambda_{D}\right) \nonumber \\
&\approx \Lambda_{D} \max\left(2, 1100\left(\frac{m_{\chi}}{\text{MeV}}\right)^{-2/15} \right. \nonumber \\
& \hphantom{\Lambda_{D}\max\left(2, 1100\right.} \left. \left(\frac{\Lambda_{D}}{1\text{ K}}\right)^{-38/15} \left(\frac{f_{D}}{0.4\%}\right)^{8/15}\right) \nonumber \\
\label{init-cond}
\end{align}

\section{Existing Limits}
As alluded to in the introduction, the constraints on composite mCPs can be quite different from elementary mCPs of the same charge. We elucidate further below. \\

\textbf{Stellar bounds:}
For $\Lambda_D \ll 1 \, \textrm{keV}$, the relevant degrees of freedom in the interior of stars and supernovae are the elementary mCPs, and their charge is restricted to $\epsilon_s < 10^{-14}$ for small enough $m_s$. The Q-balls are never produced in stellar environments. However the limit on the elementary charges translates to a limit on Q-ball charge:
\begin{equation}
\epsilon_{\chi} < 10^{-14} \frac{m_\chi}{\Lambda_D}.   
\label{stellar}
\end{equation}

\textbf{BBN and CMB } $\mathbf{N_{\rm \bf eff}:}$
As we have seen in the previous section, when there is significant thermal contact with baryons and $T_b \gg \Lambda_D$, $\mathcal{F}_{\rm Q-ball}\rightarrow 0$ and the relevant degrees of freedom are the elementary charges before recombination. Thermal equilibrium with the SM is reached only if~\cite{Munoz:2018pzp,Creque-Sarbinowski:2019mcm} $\epsilon_s \gtrsim 10^{-8} \left(\frac{m_\chi}{10~{\rm K}}\right)^\frac{1}{2}$.  This is more restrictive than stellar constraints only when $m_s\approx \Lambda_D \le 1~\mu\textrm{eV}$. Dark photons arising from bremsstrahlung and mesons from dark fusion are produced at the temperature of the dark bath and hence do not contribute appreciably to $N_{\rm eff}$ either.
\\

\textbf{CMB power spectrum:}
The effect of mCP scattering on protons was investigated in~\cite{Kovetz:2018zan}, and constraints from Planck 2015 data effectively ruled out mCPs as a solution to EDGES for $f_D>0.4\%$. It is interesting to note that since these limits only depend on the charge to mass ratio $\frac{\epsilon_\chi}{m_\chi}$, they apply equally to Q-balls as well as elementary charges. However, it was found in~\cite{Kovetz:2018zan} that no limits exist for $f_D\le 0.4\%$, so we restrict ourselves to smaller fractions. 
\section{Results}
\label{results}

\begin{figure}[htpb]
\centering
\includegraphics[width=0.95\linewidth]{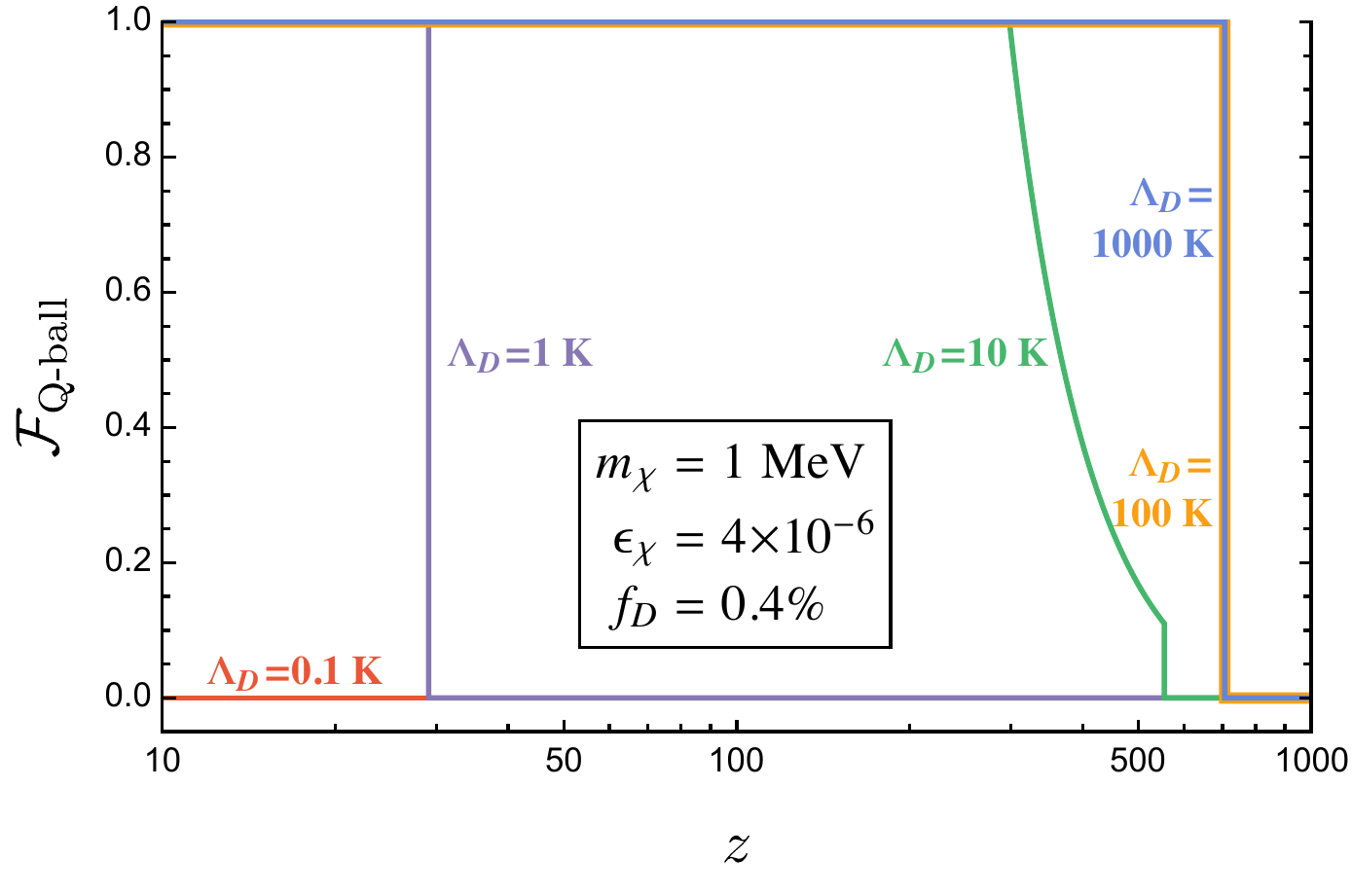}
\caption{The evolution of the fraction of the dark millicharged bath in large Q-balls}
as a function of redshift is shown for different $\Lambda_D$, the dark confining scale. Smaller $\Lambda_D$ leads to smaller Q-ball fractions.
\label{fvsz}
\end{figure}

We now display results obtained by numerically solving the coupled differential equations for time evolution. We consider a benchmark Q-ball mass $m_\chi=1$ MeV, and charge $\epsilon_\chi=4\times10^{-6}$ and $f_D=0.4\%$. We start by tracking $\mathcal{F}_{\rm Q-ball}(z)$ for different $\Lambda_D$ in Fig.~\ref{fvsz}. For large $z$, the Q-balls are broken apart into their constituents so $\mathcal{F}_{\rm Q-ball} = 0$. For lower $z$, cooling due to Hubble expansion results in a phase transition and the subsequent formation of large Q-balls, causing an increase in $\mathcal{F}_{\rm Q-ball}$. We see that for greater $\Lambda_D$, the increase in the fraction of the dark bath in Q-balls happens earlier as it is easier for Q-balls to form.

\begin{figure}[htpb]
\centering
\includegraphics[width=0.95\linewidth]{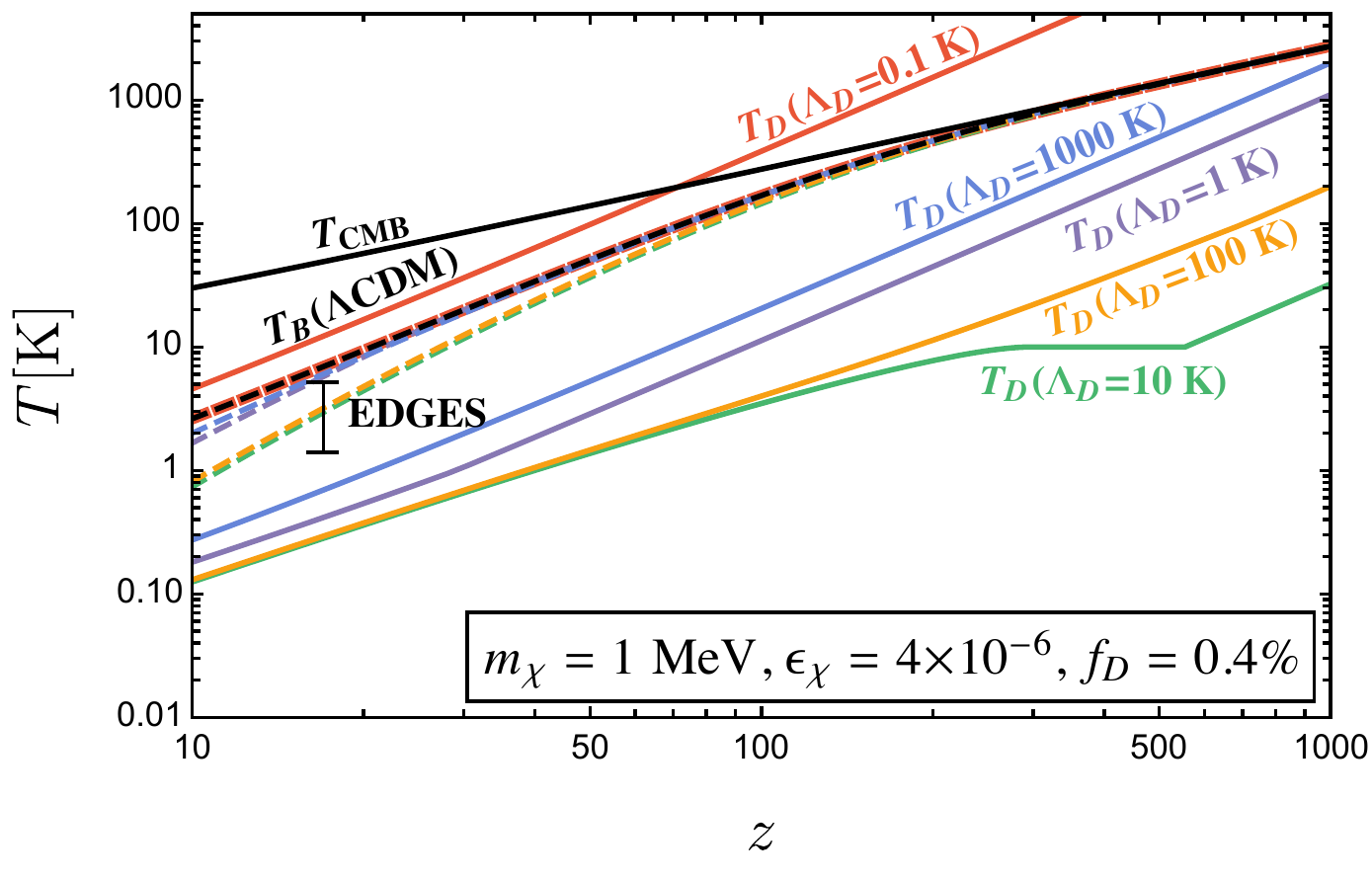}
\caption{Temperature evolution of the baryonic and DM bath are plotted as a function of redshift $z$. The CMB temperature and the baryon temperature without DM are plotted in black. The solid lines track the evolution of the dark temperature $T_D$ for different $\Lambda_D$, the dark confining temperature. The dashed lines track the baryon temperature $T_b$ for different $\Lambda_D$ with the same color code as $T_D$. The error bar marks the baryonic temperature at $z=17$ as measured by the EDGES collaboration.
For $\Lambda_D \gtrsim 10\text{ K}$ the initial dark temperature decreases with lower $\Lambda_D$, but for low enough $\Lambda_D$ the requirement from Eqn.~\ref{init-cond} becomes restrictive so the initial dark temperature begins to increase with lower $\Lambda_D$.}
\label{tvsz}
\end{figure}

In Fig.~\ref{tvsz}, the time evolution of the baryonic temperature $T_b$ and the dark temperature $T_D$ are shown for different choices of $\Lambda_D$, the dark confining scale. The CMB temperature $T_{\rm CMB}$ and the baryon temperature $T_b$ in the absence of interacting DM are shown in black for reference. 

The solid and dashed colored lines track the dark and baryonic temperatures for different $\Lambda_D$. We see that as per Eqn.~\ref{zmax-tr}, models with smaller $\Lambda_D$ are forced to reach the phase transition temperature at later times, and correspondingly must have a greater dark temperature at $z=1000$, as in Eqn.~\ref{init-cond}. For the choice of parameters in the figure, this condition is restrictive only when $\Lambda_D \lesssim 10\text{ K}$, in which case the initial dark temperature is too large for there to be much cooling of SM baryons. The requirement that the Q-balls are initially broken apart prevents models with $\Lambda_D \gtrsim 1000\text{ K}$ from being effective at cooling, through a similar increase in the initial dark temperature.

\begin{figure*}[htpb]
\centering
\includegraphics[width=0.47\linewidth]{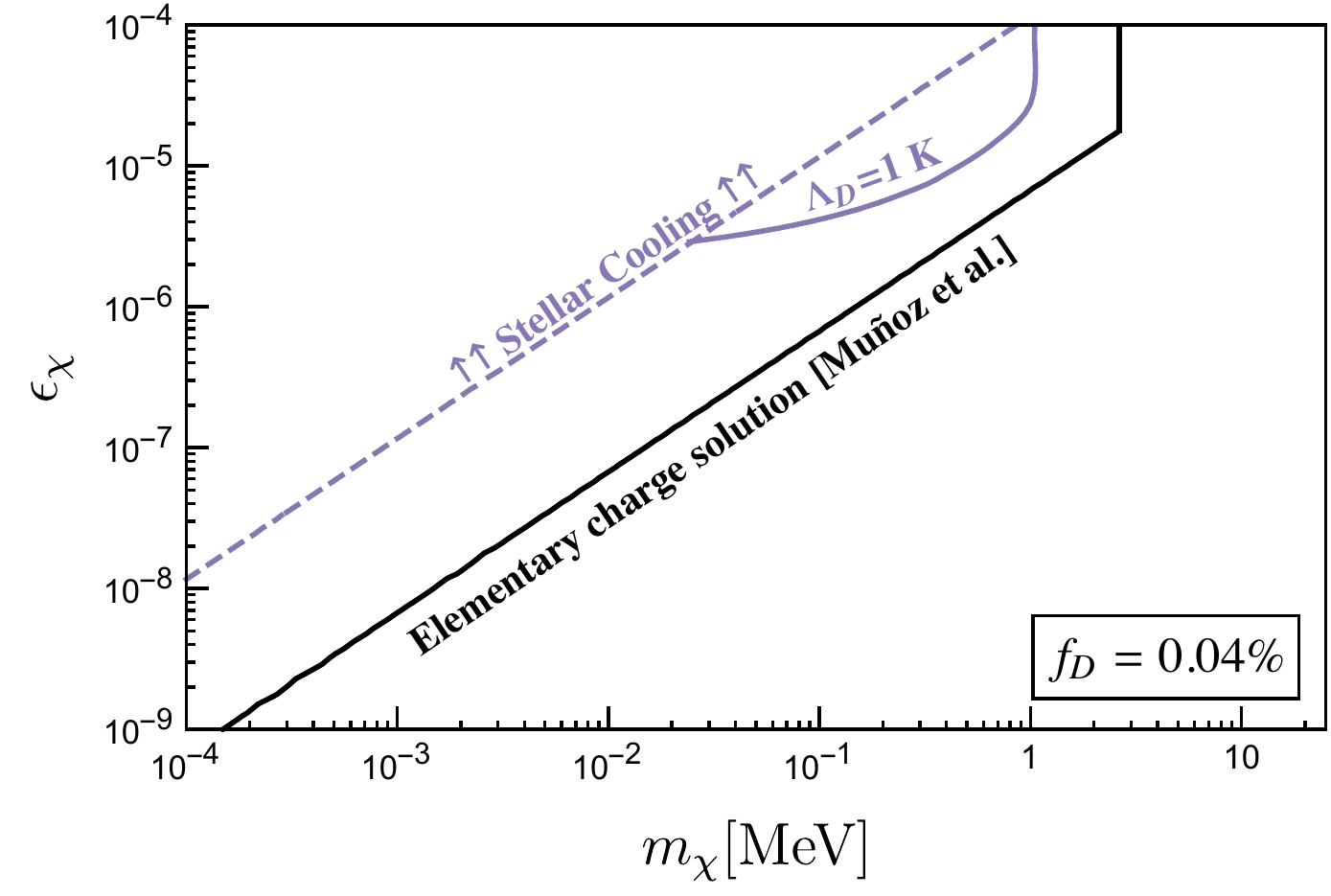}
\includegraphics[width=0.47\linewidth]{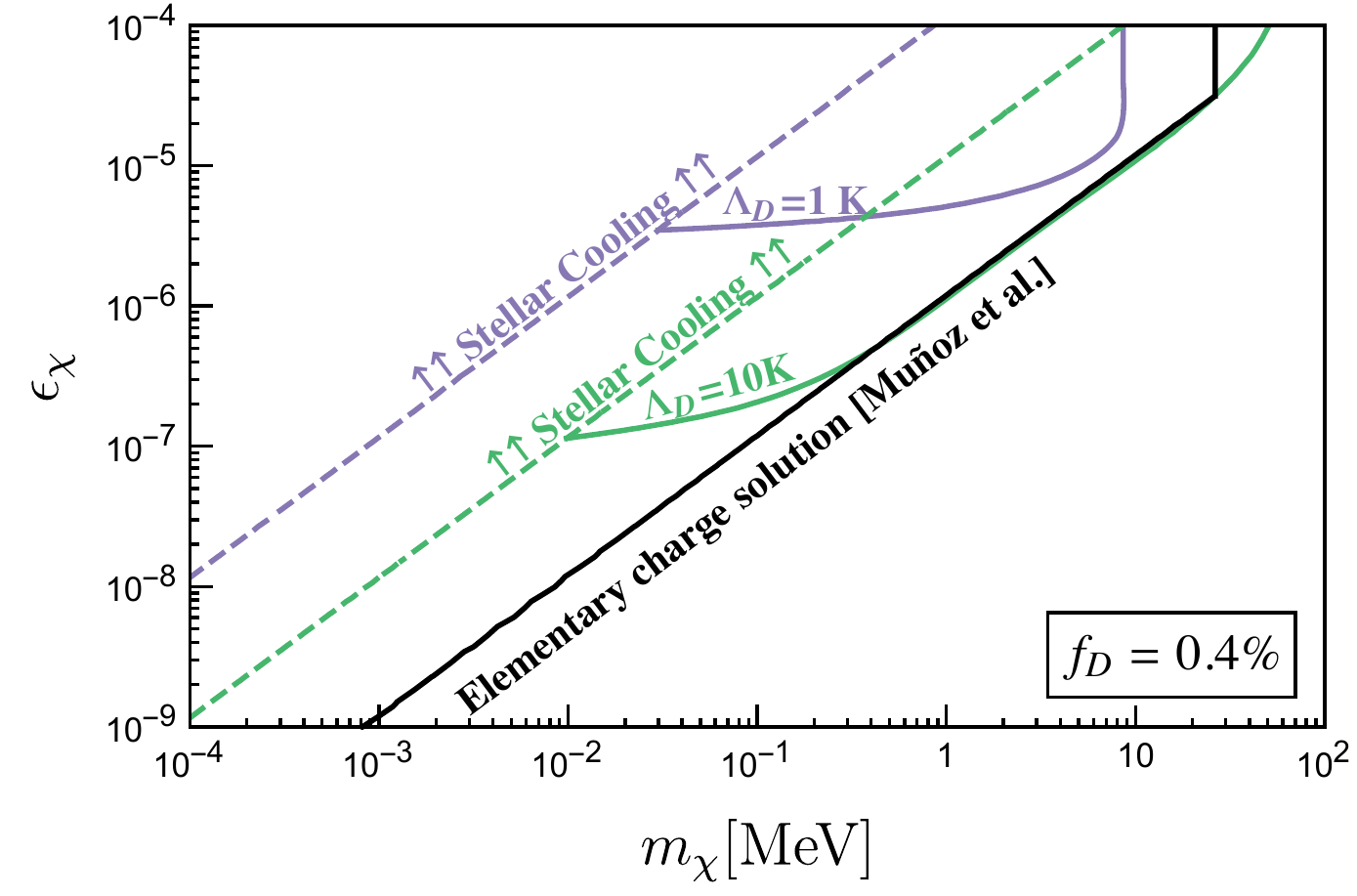}

\caption{The contours that explain the EDGES anomaly in the Q-ball charge $\epsilon_\chi$ vs Q-ball mass $m_\chi$ plane are shown for different choices of the confining scale $\Lambda_D$ for mCP bath fractions of $f_D=0.04\%$ (left) and $f_D=0.4\%$ (right). Also shown are stellar cooling constraints from Eqn.~\ref{stellar}. The elementary charge solution from \cite{Munoz:2018pzp} is shown in black.}
\label{evsm}
\end{figure*}

We next discuss the contours that explain EDGES in the $\epsilon_\chi$ vs $m_\chi$ plane and compare it to the parameter space derived for elementary charges in \cite{Munoz:2018pzp}. Given a DM fraction, elementary charges that explain EDGES obey $\epsilon_{\rm elem} \propto m_{\rm elem}$ (as seen with the black curve). This happens due to the following reason: for a fixed DM fraction, a drop in $T_b$, $\Delta T_b$ is associated with an increase in dark temperature $\Delta T_D \propto m_{\rm elem} \times \Delta T_b$, i.e. larger elementary masses $m_{\rm elem}$ undergo larger temperature gain because of equipartition. Another way to see this is that the total energy gained is equal to $n_{\rm elem} \times \Delta T_D$ and the number density is inversely proportional to $m_{\rm elem}$, and hence $T_D$ is directly proportional to $m_{\rm elem}$. Starting with an initially-cold dark bath $T_D \ll T_b$, the proportionality factor ensures that $T_D\propto m_{\rm elem}$ throughout.  This in turn implies that the elementary charges' thermal velocity is independent of the mCP mass. Finally, the heat transfer is proportional to the transfer cross-section given in Eqn.~\ref{transfer}, which is dependent only on the charge to mass ratio since the velocity is mass-independent. Thus, this behavior applies to very small masses. It was also pointed out in \cite{Munoz:2018pzp} that for a choice of DM fraction, there is also a maximum mass due to the same equipartition arguments, $m_{\rm elem} \le \mu_b f_{\rm elem} \Omega_c/\Omega_b$. The elementary charge required to explain EDGES obeys \cite{Munoz:2018pzp},

\begin{equation}
\epsilon_{\rm elem} \approx 6\times 10^{-7} \left(\frac{m_{\rm elem}}{\rm MeV}\right) \left(\frac{10^{-2}}{f_{\rm elem}}\right)^{\frac{3}{4}}.
\end{equation}
It is important to note that the entirety of the elementary charge solution is ruled out~\cite{Creque-Sarbinowski:2019mcm}. 

Next we discuss the contours for Q-balls with different $\Lambda_D$, shown in Fig.~\ref{evsm}. In each case, we mark out the unphysical region where the constituent elementary charges are ruled out by stellar constraints from Eqn.~\ref{stellar}. For the same reason as explained for the elementary charge solution, we observe a linear relationship $\epsilon_\chi \propto m_\chi$ for Q-balls in the intermediate mass range (most clearly visible for $\Lambda_D = 10\text{ K}$). However, at lower mass there is deviation from this behavior as the requirement from Eqn.~\ref{zmax-tr} becomes restrictive and forces a large initial dark temperature. This means that a greater millicharge is required for cooling than would be expected from the $\epsilon_\chi \propto m_\chi$ relation. Once again, there exists a cut-off mass that is now $\Lambda_D$ dependent. For smaller $\Lambda_D$, the mCP bath stays elementary for longer, i.e. $\mathcal{F}_{\rm Q-ball}= 0$ for longer. To compensate, a smaller Q-ball mass $m_\chi$ is required to increase heat capacity, so as to reach temperatures below $\Lambda_D$ sufficiently soon. As a corollary, larger $\Lambda_D$ results in an enhanced range in mass where the EDGES solution is viable. However, larger $\Lambda_D$ translates to stricter stellar constraints and for large enough $\Lambda_D$, the charge required to explain EDGES is ruled out by Eqn.~\ref{stellar}. The Q-balls in most of the parameter space shown in Fig.~\ref{evsm} do not survive galaxy formation. The parameter space for which Q-balls do not break up in the galaxy is given in Eqn.~\ref{gal} and can be recast as,
\begin{equation}
    m_{\chi}^{\rm gal} \lesssim 86~\textrm{eV} \left(\frac{\Lambda_D}{1~\textrm{K}}\right)
\end{equation}
Thus even for $\Lambda_D\approx 10~\textrm{K}$, the Q-balls resize themselves to masses below $1~\textrm{keV}$, making prospects for direct detection tricky.
\section{Conclusion}
Making mCDM inherently composite is a simple nuance with parallels in SM baryons. In this work, we have considered this possibility and explored its myriad consequences with specific emphasis on explaining the EDGES anomaly. The DM degrees of freedom are Q-balls at temperatures below the confining scale and elementary charges at temperatures above it. For an appropriately chosen confining scale $\Lambda_D$, the elementary charges are the degrees of freedom during BBN, CMB and in the interior of stars. The elementary charges are chosen to be feeble enough to evade all these constraints. However, at temperatures below the confining scale, these rapidly fuse into Q-balls increasing in size till they reach a size determined by stability considerations due to repulsion. These Q-balls now have large enough charges that coherently scatter with baryons at temperatures around $z=17$, relevant for physics during the dark ages, without suffering from the strict stellar and cosmology constraints that apply to elementary mCPs. Thus, we find a large unconstrained parameter space for mCP Q-balls for $f_D \le 0.4\%$, that explains the EDGES anomaly. In the next few years, this signal will also be accessible to a slew of experiments sensitive to the global 21 cm signal such as SARAS2~\cite{Singh:2017cnp}, LEDA~\cite{price2018design}, SCI-HI/PRIZM~\cite{Voytek:2013nua}, HYPERION~\cite{Presley:2015yxa} and CTP~\cite{Nhan:2016ffl}. We also find that there is a novel dark phase, where the dark bath can exist as an admixture of elementary charges that do not interact with baryons and composite Q-balls that do, with the fraction in each adjusting so as to balance the heat transfer from baryons with cooling due to Hubble expansion. This keeps the dark bath at a constant temperature until the baryons become cool enough that Hubble cooling dominates heat transfer from baryons. While this phase was an intriguing curiosity in this work, in the early universe this can have interesting consequences to thermal freeze-out of mCPs with a confining force. Finally, the avoidance of stellar and cosmology constraints due to the composite nature of the DM Q-balls provides a vastly larger parameter space that is unconstrained compared to elementary mCPs that do not confine. It is interesting to ask if these Q-balls can be probed in terrestrial experiments. This task is made more difficult by the fact that galaxy formation has the potential to destabilize the Q-balls. Galaxy formation results in DM gaining virial velocities velocities $v_{\rm vir}\approx 10^{-3}$. Self-interactions are large enough to break up the Q-balls once more if the kinetic energy exceeds the confining scale. Thus the Q-balls stay intact till today only if
\begin{equation}
m_{\chi} v_{\rm vir}^2 \lesssim \Lambda_D.
\label{gal}
\end{equation}
Thus for large enough Q-ball masses $m_\chi$, there is significant fission in galaxies, the Q-balls are resized into smaller ones that obey Eqn.~\eqref{gal} which are present in the galaxy today.  These smaller Q-balls should nevertheless be present in the galaxy today since the dark photon sets the range for self-interactions~\cite{Lasenby:2020rlf} and cuts off  long-range galactic processes such as evacuation from the galactic disk~\cite{Chuzhoy:2008zy, Dunsky:2018mqs} and retention in galactic magnetic fields~\cite{Harnik:2020ugb}, and prevents the mCP from being  blown away by the solar wind~\cite{Dunsky:2018mqs,Emken:2019tni}. 

Independent of its implications for the EDGES anomaly, this parameter space increases the scope of direct detection experiments sensitive to masses lower than $1$~MeV, albeit at momentum transfers smaller than $R_{\rm Q-ball}^{-1}$ to retain coherence. Experiments such as SENSEI~\cite{Barak:2020fql}, DAMIC~\cite{Aguilar-Arevalo:2020oii}, super-CDMS~\cite{Agnese:2017njq}, and even future proposals~\cite{Griffin:2018bjn,Essig:2019kfe, Bunting:2017net, Chen:2020jia} are not sensitive to momentum transfers $q \le \Lambda_D\approx \textrm{meV}$. Instead, manipulation with electric and magnetic fields \cite{Berlin:2019uco} is a promising detection strategy. For large enough Q-ball charge, terrestrial accumulation and subsequent detection~\cite{Pospelov:2020ktu} might be a viable avenue.

\section*{Acknowledgments}
We thank Asher Berlin, Diego Redigolo, Gordan Krnjaic and Hongwan Liu for useful discussions. SR is  supported in part by the NSF under grant PHY-1818899.   SR is also supported by the DoE under a QuantISED grant for MAGIS and the SQMS quantum center. HR is supported in part by  NSF Grant PHY-1720397 and the Gordon and Betty Moore Foundation Grant GBMF7946. 

\bibliography{biblio.bib}

\begin{thebibliography}{38}%
\makeatletter
\providecommand \@ifxundefined [1]{%
 \@ifx{#1\undefined}
}%
\providecommand \@ifnum [1]{%
 \ifnum #1\expandafter \@firstoftwo
 \else \expandafter \@secondoftwo
 \fi
}%
\providecommand \@ifx [1]{%
 \ifx #1\expandafter \@firstoftwo
 \else \expandafter \@secondoftwo
 \fi
}%
\providecommand \natexlab [1]{#1}%
\providecommand \enquote  [1]{``#1''}%
\providecommand \bibnamefont  [1]{#1}%
\providecommand \bibfnamefont [1]{#1}%
\providecommand \citenamefont [1]{#1}%
\providecommand \href@noop [0]{\@secondoftwo}%
\providecommand \href [0]{\begingroup \@sanitize@url \@href}%
\providecommand \@href[1]{\@@startlink{#1}\@@href}%
\providecommand \@@href[1]{\endgroup#1\@@endlink}%
\providecommand \@sanitize@url [0]{\catcode `\\12\catcode `\$12\catcode
  `\&12\catcode `\#12\catcode `\^12\catcode `\_12\catcode `\%12\relax}%
\providecommand \@@startlink[1]{}%
\providecommand \@@endlink[0]{}%
\providecommand \url  [0]{\begingroup\@sanitize@url \@url }%
\providecommand \@url [1]{\endgroup\@href {#1}{\urlprefix }}%
\providecommand \urlprefix  [0]{URL }%
\providecommand \Eprint [0]{\href }%
\providecommand \doibase [0]{http://dx.doi.org/}%
\providecommand \selectlanguage [0]{\@gobble}%
\providecommand \bibinfo  [0]{\@secondoftwo}%
\providecommand \bibfield  [0]{\@secondoftwo}%
\providecommand \translation [1]{[#1]}%
\providecommand \BibitemOpen [0]{}%
\providecommand \bibitemStop [0]{}%
\providecommand \bibitemNoStop [0]{.\EOS\space}%
\providecommand \EOS [0]{\spacefactor3000\relax}%
\providecommand \BibitemShut  [1]{\csname bibitem#1\endcsname}%
\let\auto@bib@innerbib\@empty
\bibitem [{\citenamefont {Bowman}\ \emph {et~al.}(2018)\citenamefont {Bowman},
  \citenamefont {Rogers}, \citenamefont {Monsalve}, \citenamefont {Mozdzen},\
  and\ \citenamefont {Mahesh}}]{Bowman:2018yin}%
  \BibitemOpen
  \bibfield  {author} {\bibinfo {author} {\bibfnamefont {Judd~D.}\ \bibnamefont
  {Bowman}}, \bibinfo {author} {\bibfnamefont {Alan E.~E.}\ \bibnamefont
  {Rogers}}, \bibinfo {author} {\bibfnamefont {Raul~A.}\ \bibnamefont
  {Monsalve}}, \bibinfo {author} {\bibfnamefont {Thomas~J.}\ \bibnamefont
  {Mozdzen}}, \ and\ \bibinfo {author} {\bibfnamefont {Nivedita}\ \bibnamefont
  {Mahesh}},\ }\bibfield  {title} {\enquote {\bibinfo {title} {{An absorption
  profile centred at 78 megahertz in the sky-averaged spectrum}},}\ }\href
  {\doibase 10.1038/nature25792} {\bibfield  {journal} {\bibinfo  {journal}
  {Nature}\ }\textbf {\bibinfo {volume} {555}},\ \bibinfo {pages} {67--70}
  (\bibinfo {year} {2018})},\ \Eprint {http://arxiv.org/abs/1810.05912}
  {arXiv:1810.05912 [astro-ph.CO]} \BibitemShut {NoStop}%
\bibitem [{\citenamefont {Barkana}(2018)}]{Barkana:2018lgd}%
  \BibitemOpen
  \bibfield  {author} {\bibinfo {author} {\bibfnamefont {Rennan}\ \bibnamefont
  {Barkana}},\ }\bibfield  {title} {\enquote {\bibinfo {title} {{Possible
  interaction between baryons and dark-matter particles revealed by the first
  stars}},}\ }\href {\doibase 10.1038/nature25791} {\bibfield  {journal}
  {\bibinfo  {journal} {Nature}\ }\textbf {\bibinfo {volume} {555}},\ \bibinfo
  {pages} {71--74} (\bibinfo {year} {2018})},\ \Eprint
  {http://arxiv.org/abs/1803.06698} {arXiv:1803.06698 [astro-ph.CO]}
  \BibitemShut {NoStop}%
\bibitem [{\citenamefont {Barkana}\ \emph {et~al.}(2018)\citenamefont
  {Barkana}, \citenamefont {Outmezguine}, \citenamefont {Redigolo},\ and\
  \citenamefont {Volansky}}]{Barkana:2018qrx}%
  \BibitemOpen
  \bibfield  {author} {\bibinfo {author} {\bibfnamefont {Rennan}\ \bibnamefont
  {Barkana}}, \bibinfo {author} {\bibfnamefont {Nadav~Joseph}\ \bibnamefont
  {Outmezguine}}, \bibinfo {author} {\bibfnamefont {Diego}\ \bibnamefont
  {Redigolo}}, \ and\ \bibinfo {author} {\bibfnamefont {Tomer}\ \bibnamefont
  {Volansky}},\ }\bibfield  {title} {\enquote {\bibinfo {title} {{Strong
  constraints on light dark matter interpretation of the EDGES signal}},}\
  }\href {\doibase 10.1103/PhysRevD.98.103005} {\bibfield  {journal} {\bibinfo
  {journal} {Phys. Rev. D}\ }\textbf {\bibinfo {volume} {98}},\ \bibinfo
  {pages} {103005} (\bibinfo {year} {2018})},\ \Eprint
  {http://arxiv.org/abs/1803.03091} {arXiv:1803.03091 [hep-ph]} \BibitemShut
  {NoStop}%
\bibitem [{\citenamefont {Kovetz}\ \emph {et~al.}(2018)\citenamefont {Kovetz},
  \citenamefont {Poulin}, \citenamefont {Gluscevic}, \citenamefont {Boddy},
  \citenamefont {Barkana},\ and\ \citenamefont
  {Kamionkowski}}]{Kovetz:2018zan}%
  \BibitemOpen
  \bibfield  {author} {\bibinfo {author} {\bibfnamefont {Ely~D.}\ \bibnamefont
  {Kovetz}}, \bibinfo {author} {\bibfnamefont {Vivian}\ \bibnamefont {Poulin}},
  \bibinfo {author} {\bibfnamefont {Vera}\ \bibnamefont {Gluscevic}}, \bibinfo
  {author} {\bibfnamefont {Kimberly~K.}\ \bibnamefont {Boddy}}, \bibinfo
  {author} {\bibfnamefont {Rennan}\ \bibnamefont {Barkana}}, \ and\ \bibinfo
  {author} {\bibfnamefont {Marc}\ \bibnamefont {Kamionkowski}},\ }\bibfield
  {title} {\enquote {\bibinfo {title} {{Tighter limits on dark matter
  explanations of the anomalous EDGES 21 cm signal}},}\ }\href {\doibase
  10.1103/PhysRevD.98.103529} {\bibfield  {journal} {\bibinfo  {journal} {Phys.
  Rev. D}\ }\textbf {\bibinfo {volume} {98}},\ \bibinfo {pages} {103529}
  (\bibinfo {year} {2018})},\ \Eprint {http://arxiv.org/abs/1807.11482}
  {arXiv:1807.11482 [astro-ph.CO]} \BibitemShut {NoStop}%
\bibitem [{\citenamefont {Berlin}\ \emph {et~al.}(2018)\citenamefont {Berlin},
  \citenamefont {Hooper}, \citenamefont {Krnjaic},\ and\ \citenamefont
  {McDermott}}]{Berlin:2018sjs}%
  \BibitemOpen
  \bibfield  {author} {\bibinfo {author} {\bibfnamefont {Asher}\ \bibnamefont
  {Berlin}}, \bibinfo {author} {\bibfnamefont {Dan}\ \bibnamefont {Hooper}},
  \bibinfo {author} {\bibfnamefont {Gordan}\ \bibnamefont {Krnjaic}}, \ and\
  \bibinfo {author} {\bibfnamefont {Samuel~D.}\ \bibnamefont {McDermott}},\
  }\bibfield  {title} {\enquote {\bibinfo {title} {{Severely Constraining Dark
  Matter Interpretations of the 21-cm Anomaly}},}\ }\href {\doibase
  10.1103/PhysRevLett.121.011102} {\bibfield  {journal} {\bibinfo  {journal}
  {Phys. Rev. Lett.}\ }\textbf {\bibinfo {volume} {121}},\ \bibinfo {pages}
  {011102} (\bibinfo {year} {2018})},\ \Eprint
  {http://arxiv.org/abs/1803.02804} {arXiv:1803.02804 [hep-ph]} \BibitemShut
  {NoStop}%
\bibitem [{\citenamefont {Creque-Sarbinowski}\ \emph
  {et~al.}(2019)\citenamefont {Creque-Sarbinowski}, \citenamefont {Ji},
  \citenamefont {Kovetz},\ and\ \citenamefont
  {Kamionkowski}}]{Creque-Sarbinowski:2019mcm}%
  \BibitemOpen
  \bibfield  {author} {\bibinfo {author} {\bibfnamefont {Cyril}\ \bibnamefont
  {Creque-Sarbinowski}}, \bibinfo {author} {\bibfnamefont {Lingyuan}\
  \bibnamefont {Ji}}, \bibinfo {author} {\bibfnamefont {Ely~D.}\ \bibnamefont
  {Kovetz}}, \ and\ \bibinfo {author} {\bibfnamefont {Marc}\ \bibnamefont
  {Kamionkowski}},\ }\bibfield  {title} {\enquote {\bibinfo {title} {{Direct
  millicharged dark matter cannot explain the EDGES signal}},}\ }\href
  {\doibase 10.1103/PhysRevD.100.023528} {\bibfield  {journal} {\bibinfo
  {journal} {Phys. Rev. D}\ }\textbf {\bibinfo {volume} {100}},\ \bibinfo
  {pages} {023528} (\bibinfo {year} {2019})},\ \Eprint
  {http://arxiv.org/abs/1903.09154} {arXiv:1903.09154 [astro-ph.CO]}
  \BibitemShut {NoStop}%
\bibitem [{\citenamefont {Feng}\ and\ \citenamefont
  {Holder}(2018)}]{Feng:2018rje}%
  \BibitemOpen
  \bibfield  {author} {\bibinfo {author} {\bibfnamefont {Chang}\ \bibnamefont
  {Feng}}\ and\ \bibinfo {author} {\bibfnamefont {Gilbert}\ \bibnamefont
  {Holder}},\ }\bibfield  {title} {\enquote {\bibinfo {title} {{Enhanced global
  signal of neutral hydrogen due to excess radiation at cosmic dawn}},}\ }\href
  {\doibase 10.3847/2041-8213/aac0fe} {\bibfield  {journal} {\bibinfo
  {journal} {Astrophys. J. Lett.}\ }\textbf {\bibinfo {volume} {858}},\
  \bibinfo {pages} {L17} (\bibinfo {year} {2018})},\ \Eprint
  {http://arxiv.org/abs/1802.07432} {arXiv:1802.07432 [astro-ph.CO]}
  \BibitemShut {NoStop}%
\bibitem [{\citenamefont {Ewall-Wice}\ \emph {et~al.}(2018)\citenamefont
  {Ewall-Wice}, \citenamefont {Chang}, \citenamefont {Lazio}, \citenamefont
  {Dore}, \citenamefont {Seiffert},\ and\ \citenamefont
  {Monsalve}}]{Ewall-Wice:2018bzf}%
  \BibitemOpen
  \bibfield  {author} {\bibinfo {author} {\bibfnamefont {A.}~\bibnamefont
  {Ewall-Wice}}, \bibinfo {author} {\bibfnamefont {T.~C.}\ \bibnamefont
  {Chang}}, \bibinfo {author} {\bibfnamefont {J.}~\bibnamefont {Lazio}},
  \bibinfo {author} {\bibfnamefont {O.}~\bibnamefont {Dore}}, \bibinfo {author}
  {\bibfnamefont {M.}~\bibnamefont {Seiffert}}, \ and\ \bibinfo {author}
  {\bibfnamefont {R.~A.}\ \bibnamefont {Monsalve}},\ }\bibfield  {title}
  {\enquote {\bibinfo {title} {{Modeling the Radio Background from the First
  Black Holes at Cosmic Dawn: Implications for the 21 cm Absorption
  Amplitude}},}\ }\href {\doibase 10.3847/1538-4357/aae51d} {\bibfield
  {journal} {\bibinfo  {journal} {Astrophys. J.}\ }\textbf {\bibinfo {volume}
  {868}},\ \bibinfo {pages} {63} (\bibinfo {year} {2018})},\ \Eprint
  {http://arxiv.org/abs/1803.01815} {arXiv:1803.01815 [astro-ph.CO]}
  \BibitemShut {NoStop}%
\bibitem [{\citenamefont {Pospelov}\ \emph {et~al.}(2018)\citenamefont
  {Pospelov}, \citenamefont {Pradler}, \citenamefont {Ruderman},\ and\
  \citenamefont {Urbano}}]{Pospelov:2018kdh}%
  \BibitemOpen
  \bibfield  {author} {\bibinfo {author} {\bibfnamefont {Maxim}\ \bibnamefont
  {Pospelov}}, \bibinfo {author} {\bibfnamefont {Josef}\ \bibnamefont
  {Pradler}}, \bibinfo {author} {\bibfnamefont {Joshua~T.}\ \bibnamefont
  {Ruderman}}, \ and\ \bibinfo {author} {\bibfnamefont {Alfredo}\ \bibnamefont
  {Urbano}},\ }\bibfield  {title} {\enquote {\bibinfo {title} {{Room for New
  Physics in the Rayleigh-Jeans Tail of the Cosmic Microwave Background}},}\
  }\href {\doibase 10.1103/PhysRevLett.121.031103} {\bibfield  {journal}
  {\bibinfo  {journal} {Phys. Rev. Lett.}\ }\textbf {\bibinfo {volume} {121}},\
  \bibinfo {pages} {031103} (\bibinfo {year} {2018})},\ \Eprint
  {http://arxiv.org/abs/1803.07048} {arXiv:1803.07048 [hep-ph]} \BibitemShut
  {NoStop}%
\bibitem [{\citenamefont {Fialkov}\ and\ \citenamefont
  {Barkana}(2019)}]{Fialkov:2019vnb}%
  \BibitemOpen
  \bibfield  {author} {\bibinfo {author} {\bibfnamefont {Anastasia}\
  \bibnamefont {Fialkov}}\ and\ \bibinfo {author} {\bibfnamefont {Rennan}\
  \bibnamefont {Barkana}},\ }\bibfield  {title} {\enquote {\bibinfo {title}
  {{Signature of Excess Radio Background in the 21-cm Global Signal and Power
  Spectrum}},}\ }\href {\doibase 10.1093/mnras/stz873} {\bibfield  {journal}
  {\bibinfo  {journal} {Mon. Not. Roy. Astron. Soc.}\ }\textbf {\bibinfo
  {volume} {486}},\ \bibinfo {pages} {1763--1773} (\bibinfo {year} {2019})},\
  \Eprint {http://arxiv.org/abs/1902.02438} {arXiv:1902.02438 [astro-ph.CO]}
  \BibitemShut {NoStop}%
\bibitem [{\citenamefont {Liu}\ \emph {et~al.}(2019)\citenamefont {Liu},
  \citenamefont {Outmezguine}, \citenamefont {Redigolo},\ and\ \citenamefont
  {Volansky}}]{Liu:2019knx}%
  \BibitemOpen
  \bibfield  {author} {\bibinfo {author} {\bibfnamefont {Hongwan}\ \bibnamefont
  {Liu}}, \bibinfo {author} {\bibfnamefont {Nadav~Joseph}\ \bibnamefont
  {Outmezguine}}, \bibinfo {author} {\bibfnamefont {Diego}\ \bibnamefont
  {Redigolo}}, \ and\ \bibinfo {author} {\bibfnamefont {Tomer}\ \bibnamefont
  {Volansky}},\ }\bibfield  {title} {\enquote {\bibinfo {title} {{Reviving
  Millicharged Dark Matter for 21-cm Cosmology}},}\ }\href {\doibase
  10.1103/PhysRevD.100.123011} {\bibfield  {journal} {\bibinfo  {journal}
  {Phys. Rev. D}\ }\textbf {\bibinfo {volume} {100}},\ \bibinfo {pages}
  {123011} (\bibinfo {year} {2019})},\ \Eprint
  {http://arxiv.org/abs/1908.06986} {arXiv:1908.06986 [hep-ph]} \BibitemShut
  {NoStop}%
\bibitem [{\citenamefont {Grabowska}\ \emph {et~al.}(2018)\citenamefont
  {Grabowska}, \citenamefont {Melia},\ and\ \citenamefont
  {Rajendran}}]{Grabowska:2018lnd}%
  \BibitemOpen
  \bibfield  {author} {\bibinfo {author} {\bibfnamefont {Dorota~M.}\
  \bibnamefont {Grabowska}}, \bibinfo {author} {\bibfnamefont {Tom}\
  \bibnamefont {Melia}}, \ and\ \bibinfo {author} {\bibfnamefont {Surjeet}\
  \bibnamefont {Rajendran}},\ }\bibfield  {title} {\enquote {\bibinfo {title}
  {{Detecting Dark Blobs}},}\ }\href {\doibase 10.1103/PhysRevD.98.115020}
  {\bibfield  {journal} {\bibinfo  {journal} {Phys. Rev. D}\ }\textbf {\bibinfo
  {volume} {98}},\ \bibinfo {pages} {115020} (\bibinfo {year} {2018})},\
  \Eprint {http://arxiv.org/abs/1807.03788} {arXiv:1807.03788 [hep-ph]}
  \BibitemShut {NoStop}%
\bibitem [{\citenamefont {Kusenko}\ and\ \citenamefont
  {Shaposhnikov}(1998)}]{Kusenko}%
  \BibitemOpen
  \bibfield  {author} {\bibinfo {author} {\bibfnamefont {Alexander}\
  \bibnamefont {Kusenko}}\ and\ \bibinfo {author} {\bibfnamefont {Mikhail}\
  \bibnamefont {Shaposhnikov}},\ }\bibfield  {title} {\enquote {\bibinfo
  {title} {{Supersymmetric Q-balls as dark matter}},}\ }\href {\doibase
  10.1016/S0370-2693(97)01375-0} {\bibfield  {journal} {\bibinfo  {journal}
  {Phys. Lett. B}\ }\textbf {\bibinfo {volume} {418}},\ \bibinfo {pages}
  {46--54} (\bibinfo {year} {1998})},\ \Eprint {http://arxiv.org/abs/9709492}
  {arXiv:9709492 [hep-ph]} \BibitemShut {NoStop}%
\bibitem [{\citenamefont {Krnjaic}\ and\ \citenamefont
  {Sigurdson}(2015)}]{Krnjaic:2014xza}%
  \BibitemOpen
  \bibfield  {author} {\bibinfo {author} {\bibfnamefont {Gordan}\ \bibnamefont
  {Krnjaic}}\ and\ \bibinfo {author} {\bibfnamefont {Kris}\ \bibnamefont
  {Sigurdson}},\ }\bibfield  {title} {\enquote {\bibinfo {title} {{Big Bang
  Darkleosynthesis}},}\ }\href {\doibase 10.1016/j.physletb.2015.11.001}
  {\bibfield  {journal} {\bibinfo  {journal} {Phys. Lett. B}\ }\textbf
  {\bibinfo {volume} {751}},\ \bibinfo {pages} {464--468} (\bibinfo {year}
  {2015})},\ \Eprint {http://arxiv.org/abs/1406.1171} {arXiv:1406.1171
  [hep-ph]} \BibitemShut {NoStop}%
\bibitem [{\citenamefont {Gamow}(1963)}]{gamow1963quantum}%
  \BibitemOpen
  \bibfield  {author} {\bibinfo {author} {\bibfnamefont {George}\ \bibnamefont
  {Gamow}},\ }\href@noop {} {\emph {\bibinfo {title} {The Quantum Theory of the
  Atomic Nucleus}}}\ (\bibinfo  {publisher} {US Atomic Energy Commission,
  Division of Technical Information Extension},\ \bibinfo {year}
  {1963})\BibitemShut {NoStop}%
\bibitem [{\citenamefont {Dvorkin}\ \emph {et~al.}(2014)\citenamefont
  {Dvorkin}, \citenamefont {Blum},\ and\ \citenamefont
  {Kamionkowski}}]{Dvorkin:2013cea}%
  \BibitemOpen
  \bibfield  {author} {\bibinfo {author} {\bibfnamefont {Cora}\ \bibnamefont
  {Dvorkin}}, \bibinfo {author} {\bibfnamefont {Kfir}\ \bibnamefont {Blum}}, \
  and\ \bibinfo {author} {\bibfnamefont {Marc}\ \bibnamefont {Kamionkowski}},\
  }\bibfield  {title} {\enquote {\bibinfo {title} {{Constraining Dark
  Matter-Baryon Scattering with Linear Cosmology}},}\ }\href {\doibase
  10.1103/PhysRevD.89.023519} {\bibfield  {journal} {\bibinfo  {journal} {Phys.
  Rev. D}\ }\textbf {\bibinfo {volume} {89}},\ \bibinfo {pages} {023519}
  (\bibinfo {year} {2014})},\ \Eprint {http://arxiv.org/abs/1311.2937}
  {arXiv:1311.2937 [astro-ph.CO]} \BibitemShut {NoStop}%
\bibitem [{\citenamefont {Ali-Ha\"{\i}moud}\ and\ \citenamefont
  {Hirata}(2011)}]{Hyrec1}%
  \BibitemOpen
  \bibfield  {author} {\bibinfo {author} {\bibfnamefont {Yacine}\ \bibnamefont
  {Ali-Ha\"{\i}moud}}\ and\ \bibinfo {author} {\bibfnamefont {Christopher~M.}\
  \bibnamefont {Hirata}},\ }\bibfield  {title} {\enquote {\bibinfo {title}
  {Hyrec: A fast and highly accurate primordial hydrogen and helium
  recombination code},}\ }\href {\doibase 10.1103/PhysRevD.83.043513}
  {\bibfield  {journal} {\bibinfo  {journal} {Phys. Rev. D}\ }\textbf {\bibinfo
  {volume} {83}},\ \bibinfo {pages} {043513} (\bibinfo {year} {2011})},\
  \Eprint {http://arxiv.org/abs/1011.3758} {arXiv:1011.3758 [astro-ph.CO]}
  \BibitemShut {NoStop}%
\bibitem [{\citenamefont {Lee}\ and\ \citenamefont
  {Ali-Ha\"{\i}moud}(2020)}]{Hyrec2}%
  \BibitemOpen
  \bibfield  {author} {\bibinfo {author} {\bibfnamefont {Nanoom}\ \bibnamefont
  {Lee}}\ and\ \bibinfo {author} {\bibfnamefont {Yacine}\ \bibnamefont
  {Ali-Ha\"{\i}moud}},\ }\bibfield  {title} {\enquote {\bibinfo {title}
  {Hyrec-2: A highly accurate sub-millisecond recombination code},}\ }\href
  {\doibase 10.1103/PhysRevD.102.083517} {\bibfield  {journal} {\bibinfo
  {journal} {Phys. Rev. D}\ }\textbf {\bibinfo {volume} {102}},\ \bibinfo
  {pages} {083517} (\bibinfo {year} {2020})},\ \Eprint
  {http://arxiv.org/abs/2007.14114} {arXiv:2007.14114 [astro-ph.CO]}
  \BibitemShut {NoStop}%
\bibitem [{\citenamefont {Mu\~noz}\ and\ \citenamefont
  {Loeb}(2018)}]{Munoz:2018pzp}%
  \BibitemOpen
  \bibfield  {author} {\bibinfo {author} {\bibfnamefont {Julian~B.}\
  \bibnamefont {Mu\~noz}}\ and\ \bibinfo {author} {\bibfnamefont {Abraham}\
  \bibnamefont {Loeb}},\ }\bibfield  {title} {\enquote {\bibinfo {title} {{A
  small amount of mini-charged dark matter could cool the baryons in the early
  Universe}},}\ }\href {\doibase 10.1038/s41586-018-0151-x} {\bibfield
  {journal} {\bibinfo  {journal} {Nature}\ }\textbf {\bibinfo {volume} {557}},\
  \bibinfo {pages} {684} (\bibinfo {year} {2018})},\ \Eprint
  {http://arxiv.org/abs/1802.10094} {arXiv:1802.10094 [astro-ph.CO]}
  \BibitemShut {NoStop}%
\bibitem [{\citenamefont {Singh}\ \emph {et~al.}(2018)\citenamefont {Singh}
  \emph {et~al.}}]{Singh:2017cnp}%
  \BibitemOpen
  \bibfield  {author} {\bibinfo {author} {\bibfnamefont {Saurabh}\ \bibnamefont
  {Singh}} \emph {et~al.},\ }\bibfield  {title} {\enquote {\bibinfo {title}
  {{SARAS 2 constraints on global 21-cm signals from the Epoch of
  Reionization}},}\ }\href {\doibase 10.3847/1538-4357/aabae1} {\bibfield
  {journal} {\bibinfo  {journal} {Astrophys. J.}\ }\textbf {\bibinfo {volume}
  {858}},\ \bibinfo {pages} {54} (\bibinfo {year} {2018})},\ \Eprint
  {http://arxiv.org/abs/1711.11281} {arXiv:1711.11281 [astro-ph.CO]}
  \BibitemShut {NoStop}%
\bibitem [{\citenamefont {Price}\ \emph {et~al.}(2018)\citenamefont {Price},
  \citenamefont {Greenhill}, \citenamefont {Fialkov}, \citenamefont {Bernardi},
  \citenamefont {Garsden}, \citenamefont {Barsdell}, \citenamefont {Kocz},
  \citenamefont {Anderson}, \citenamefont {Bourke}, \citenamefont {Craig} \emph
  {et~al.}}]{price2018design}%
  \BibitemOpen
  \bibfield  {author} {\bibinfo {author} {\bibfnamefont {DC}~\bibnamefont
  {Price}}, \bibinfo {author} {\bibfnamefont {LJ}~\bibnamefont {Greenhill}},
  \bibinfo {author} {\bibfnamefont {Anastasia}\ \bibnamefont {Fialkov}},
  \bibinfo {author} {\bibfnamefont {GIANNI}\ \bibnamefont {Bernardi}}, \bibinfo
  {author} {\bibfnamefont {H}~\bibnamefont {Garsden}}, \bibinfo {author}
  {\bibfnamefont {BR}~\bibnamefont {Barsdell}}, \bibinfo {author}
  {\bibfnamefont {J}~\bibnamefont {Kocz}}, \bibinfo {author} {\bibfnamefont
  {MM}~\bibnamefont {Anderson}}, \bibinfo {author} {\bibfnamefont
  {SA}~\bibnamefont {Bourke}}, \bibinfo {author} {\bibfnamefont
  {J}~\bibnamefont {Craig}},  \emph {et~al.},\ }\bibfield  {title} {\enquote
  {\bibinfo {title} {Design and characterization of the large-aperture
  experiment to detect the dark age (leda) radiometer systems},}\ }\href@noop
  {} {\bibfield  {journal} {\bibinfo  {journal} {Monthly Notices of the Royal
  Astronomical Society}\ }\textbf {\bibinfo {volume} {478}},\ \bibinfo {pages}
  {4193--4213} (\bibinfo {year} {2018})}\BibitemShut {NoStop}%
\bibitem [{\citenamefont {Voytek}\ \emph {et~al.}(2014)\citenamefont {Voytek},
  \citenamefont {Natarajan}, \citenamefont {J\'auregui~Garc\'\i{}a},
  \citenamefont {Peterson},\ and\ \citenamefont
  {L\'opez-Cruz}}]{Voytek:2013nua}%
  \BibitemOpen
  \bibfield  {author} {\bibinfo {author} {\bibfnamefont {Tabitha~C.}\
  \bibnamefont {Voytek}}, \bibinfo {author} {\bibfnamefont {Aravind}\
  \bibnamefont {Natarajan}}, \bibinfo {author} {\bibfnamefont {Jos\'e~Miguel}\
  \bibnamefont {J\'auregui~Garc\'\i{}a}}, \bibinfo {author} {\bibfnamefont
  {Jeffrey~B.}\ \bibnamefont {Peterson}}, \ and\ \bibinfo {author}
  {\bibfnamefont {Omar}\ \bibnamefont {L\'opez-Cruz}},\ }\bibfield  {title}
  {\enquote {\bibinfo {title} {{Probing the Dark Ages at $z \sim$ 20: The
  SCI-HI 21 cm All-sky Spectrum Experiment}},}\ }\href {\doibase
  10.1088/2041-8205/782/1/L9} {\bibfield  {journal} {\bibinfo  {journal}
  {Astrophys. J. Lett.}\ }\textbf {\bibinfo {volume} {782}},\ \bibinfo {pages}
  {L9} (\bibinfo {year} {2014})},\ \Eprint {http://arxiv.org/abs/1311.0014}
  {arXiv:1311.0014 [astro-ph.CO]} \BibitemShut {NoStop}%
\bibitem [{\citenamefont {Presley}\ \emph {et~al.}(2015)\citenamefont
  {Presley}, \citenamefont {Liu},\ and\ \citenamefont
  {Parsons}}]{Presley:2015yxa}%
  \BibitemOpen
  \bibfield  {author} {\bibinfo {author} {\bibfnamefont {Morgan~E.}\
  \bibnamefont {Presley}}, \bibinfo {author} {\bibfnamefont {Adrian}\
  \bibnamefont {Liu}}, \ and\ \bibinfo {author} {\bibfnamefont {Aaron~R.}\
  \bibnamefont {Parsons}},\ }\bibfield  {title} {\enquote {\bibinfo {title}
  {{Measuring the Cosmological 21 cm Monopole with an Interferometer}},}\
  }\href {\doibase 10.1088/0004-637X/809/1/18} {\bibfield  {journal} {\bibinfo
  {journal} {Astrophys. J.}\ }\textbf {\bibinfo {volume} {809}},\ \bibinfo
  {pages} {18} (\bibinfo {year} {2015})},\ \Eprint
  {http://arxiv.org/abs/1501.01633} {arXiv:1501.01633 [astro-ph.CO]}
  \BibitemShut {NoStop}%
\bibitem [{\citenamefont {Nhan}\ \emph {et~al.}(2017)\citenamefont {Nhan},
  \citenamefont {Bradley},\ and\ \citenamefont {Burns}}]{Nhan:2016ffl}%
  \BibitemOpen
  \bibfield  {author} {\bibinfo {author} {\bibfnamefont {Bang~D.}\ \bibnamefont
  {Nhan}}, \bibinfo {author} {\bibfnamefont {Richard~F.}\ \bibnamefont
  {Bradley}}, \ and\ \bibinfo {author} {\bibfnamefont {Jack~O.}\ \bibnamefont
  {Burns}},\ }\bibfield  {title} {\enquote {\bibinfo {title} {{A polarimetric
  approach for constraining the dynamic foreground spectrum for cosmological
  global 21-cm measurements}},}\ }\href {\doibase 10.3847/1538-4357/836/1/90}
  {\bibfield  {journal} {\bibinfo  {journal} {Astrophys. J.}\ }\textbf
  {\bibinfo {volume} {836}},\ \bibinfo {pages} {90} (\bibinfo {year} {2017})},\
  \Eprint {http://arxiv.org/abs/1611.06062} {arXiv:1611.06062 [astro-ph.IM]}
  \BibitemShut {NoStop}%
\bibitem [{\citenamefont {Lasenby}(2020)}]{Lasenby:2020rlf}%
  \BibitemOpen
  \bibfield  {author} {\bibinfo {author} {\bibfnamefont {Robert}\ \bibnamefont
  {Lasenby}},\ }\bibfield  {title} {\enquote {\bibinfo {title} {{Long range
  dark matter self-interactions and plasma instabilities}},}\ }\href {\doibase
  10.1088/1475-7516/2020/11/034} {\bibfield  {journal} {\bibinfo  {journal}
  {JCAP}\ }\textbf {\bibinfo {volume} {11}},\ \bibinfo {pages} {034} (\bibinfo
  {year} {2020})},\ \Eprint {http://arxiv.org/abs/2007.00667} {arXiv:2007.00667
  [hep-ph]} \BibitemShut {NoStop}%
\bibitem [{\citenamefont {Chuzhoy}\ and\ \citenamefont
  {Kolb}(2009)}]{Chuzhoy:2008zy}%
  \BibitemOpen
  \bibfield  {author} {\bibinfo {author} {\bibfnamefont {Leonid}\ \bibnamefont
  {Chuzhoy}}\ and\ \bibinfo {author} {\bibfnamefont {Edward~W.}\ \bibnamefont
  {Kolb}},\ }\bibfield  {title} {\enquote {\bibinfo {title} {{Reopening the
  window on charged dark matter}},}\ }\href {\doibase
  10.1088/1475-7516/2009/07/014} {\bibfield  {journal} {\bibinfo  {journal}
  {JCAP}\ }\textbf {\bibinfo {volume} {07}},\ \bibinfo {pages} {014} (\bibinfo
  {year} {2009})},\ \Eprint {http://arxiv.org/abs/0809.0436} {arXiv:0809.0436
  [astro-ph]} \BibitemShut {NoStop}%
\bibitem [{\citenamefont {Dunsky}\ \emph {et~al.}(2019)\citenamefont {Dunsky},
  \citenamefont {Hall},\ and\ \citenamefont {Harigaya}}]{Dunsky:2018mqs}%
  \BibitemOpen
  \bibfield  {author} {\bibinfo {author} {\bibfnamefont {David}\ \bibnamefont
  {Dunsky}}, \bibinfo {author} {\bibfnamefont {Lawrence~J.}\ \bibnamefont
  {Hall}}, \ and\ \bibinfo {author} {\bibfnamefont {Keisuke}\ \bibnamefont
  {Harigaya}},\ }\bibfield  {title} {\enquote {\bibinfo {title} {{CHAMP Cosmic
  Rays}},}\ }\href {\doibase 10.1088/1475-7516/2019/07/015} {\bibfield
  {journal} {\bibinfo  {journal} {JCAP}\ }\textbf {\bibinfo {volume} {07}},\
  \bibinfo {pages} {015} (\bibinfo {year} {2019})},\ \Eprint
  {http://arxiv.org/abs/1812.11116} {arXiv:1812.11116 [astro-ph.HE]}
  \BibitemShut {NoStop}%
\bibitem [{\citenamefont {Harnik}\ \emph {et~al.}(2020)\citenamefont {Harnik},
  \citenamefont {Plestid}, \citenamefont {Pospelov},\ and\ \citenamefont
  {Ramani}}]{Harnik:2020ugb}%
  \BibitemOpen
  \bibfield  {author} {\bibinfo {author} {\bibfnamefont {Roni}\ \bibnamefont
  {Harnik}}, \bibinfo {author} {\bibfnamefont {Ryan}\ \bibnamefont {Plestid}},
  \bibinfo {author} {\bibfnamefont {Maxim}\ \bibnamefont {Pospelov}}, \ and\
  \bibinfo {author} {\bibfnamefont {Harikrishnan}\ \bibnamefont {Ramani}},\
  }\bibfield  {title} {\enquote {\bibinfo {title} {{Millicharged Cosmic Rays
  and Low Recoil Detectors}},}\ }\href@noop {} {\  (\bibinfo {year} {2020})},\
  \Eprint {http://arxiv.org/abs/2010.11190} {arXiv:2010.11190 [hep-ph]}
  \BibitemShut {NoStop}%
\bibitem [{\citenamefont {Emken}\ \emph {et~al.}(2019)\citenamefont {Emken},
  \citenamefont {Essig}, \citenamefont {Kouvaris},\ and\ \citenamefont
  {Sholapurkar}}]{Emken:2019tni}%
  \BibitemOpen
  \bibfield  {author} {\bibinfo {author} {\bibfnamefont {Timon}\ \bibnamefont
  {Emken}}, \bibinfo {author} {\bibfnamefont {Rouven}\ \bibnamefont {Essig}},
  \bibinfo {author} {\bibfnamefont {Chris}\ \bibnamefont {Kouvaris}}, \ and\
  \bibinfo {author} {\bibfnamefont {Mukul}\ \bibnamefont {Sholapurkar}},\
  }\bibfield  {title} {\enquote {\bibinfo {title} {{Direct Detection of
  Strongly Interacting Sub-GeV Dark Matter via Electron Recoils}},}\ }\href
  {\doibase 10.1088/1475-7516/2019/09/070} {\bibfield  {journal} {\bibinfo
  {journal} {JCAP}\ }\textbf {\bibinfo {volume} {09}},\ \bibinfo {pages} {070}
  (\bibinfo {year} {2019})},\ \Eprint {http://arxiv.org/abs/1905.06348}
  {arXiv:1905.06348 [hep-ph]} \BibitemShut {NoStop}%
\bibitem [{\citenamefont {Barak}\ \emph {et~al.}(2020)\citenamefont {Barak}
  \emph {et~al.}}]{Barak:2020fql}%
  \BibitemOpen
  \bibfield  {author} {\bibinfo {author} {\bibfnamefont {Liron}\ \bibnamefont
  {Barak}} \emph {et~al.} (\bibinfo {collaboration} {SENSEI}),\ }\bibfield
  {title} {\enquote {\bibinfo {title} {{SENSEI: Direct-Detection Results on
  sub-GeV Dark Matter from a New Skipper-CCD}},}\ }\href {\doibase
  10.1103/PhysRevLett.125.171802} {\bibfield  {journal} {\bibinfo  {journal}
  {Phys. Rev. Lett.}\ }\textbf {\bibinfo {volume} {125}},\ \bibinfo {pages}
  {171802} (\bibinfo {year} {2020})},\ \Eprint
  {http://arxiv.org/abs/2004.11378} {arXiv:2004.11378 [astro-ph.CO]}
  \BibitemShut {NoStop}%
\bibitem [{\citenamefont {Aguilar-Arevalo}\ \emph {et~al.}(2020)\citenamefont
  {Aguilar-Arevalo} \emph {et~al.}}]{Aguilar-Arevalo:2020oii}%
  \BibitemOpen
  \bibfield  {author} {\bibinfo {author} {\bibfnamefont {A.}~\bibnamefont
  {Aguilar-Arevalo}} \emph {et~al.} (\bibinfo {collaboration} {DAMIC}),\
  }\bibfield  {title} {\enquote {\bibinfo {title} {{Results on low-mass weakly
  interacting massive particles from a 11 kg-day target exposure of DAMIC at
  SNOLAB}},}\ }\href {\doibase 10.1103/PhysRevLett.125.241803} {\bibfield
  {journal} {\bibinfo  {journal} {Phys. Rev. Lett.}\ }\textbf {\bibinfo
  {volume} {125}},\ \bibinfo {pages} {241803} (\bibinfo {year} {2020})},\
  \Eprint {http://arxiv.org/abs/2007.15622} {arXiv:2007.15622 [astro-ph.CO]}
  \BibitemShut {NoStop}%
\bibitem [{\citenamefont {Agnese}\ \emph {et~al.}(2018)\citenamefont {Agnese}
  \emph {et~al.}}]{Agnese:2017njq}%
  \BibitemOpen
  \bibfield  {author} {\bibinfo {author} {\bibfnamefont {R.}~\bibnamefont
  {Agnese}} \emph {et~al.} (\bibinfo {collaboration} {SuperCDMS}),\ }\bibfield
  {title} {\enquote {\bibinfo {title} {{Results from the Super Cryogenic Dark
  Matter Search Experiment at Soudan}},}\ }\href {\doibase
  10.1103/PhysRevLett.120.061802} {\bibfield  {journal} {\bibinfo  {journal}
  {Phys. Rev. Lett.}\ }\textbf {\bibinfo {volume} {120}},\ \bibinfo {pages}
  {061802} (\bibinfo {year} {2018})},\ \Eprint
  {http://arxiv.org/abs/1708.08869} {arXiv:1708.08869 [hep-ex]} \BibitemShut
  {NoStop}%
\bibitem [{\citenamefont {Griffin}\ \emph {et~al.}(2018)\citenamefont
  {Griffin}, \citenamefont {Knapen}, \citenamefont {Lin},\ and\ \citenamefont
  {Zurek}}]{Griffin:2018bjn}%
  \BibitemOpen
  \bibfield  {author} {\bibinfo {author} {\bibfnamefont {Sinead}\ \bibnamefont
  {Griffin}}, \bibinfo {author} {\bibfnamefont {Simon}\ \bibnamefont {Knapen}},
  \bibinfo {author} {\bibfnamefont {Tongyan}\ \bibnamefont {Lin}}, \ and\
  \bibinfo {author} {\bibfnamefont {Kathryn~M.}\ \bibnamefont {Zurek}},\
  }\bibfield  {title} {\enquote {\bibinfo {title} {{Directional Detection of
  Light Dark Matter with Polar Materials}},}\ }\href {\doibase
  10.1103/PhysRevD.98.115034} {\bibfield  {journal} {\bibinfo  {journal} {Phys.
  Rev. D}\ }\textbf {\bibinfo {volume} {98}},\ \bibinfo {pages} {115034}
  (\bibinfo {year} {2018})},\ \Eprint {http://arxiv.org/abs/1807.10291}
  {arXiv:1807.10291 [hep-ph]} \BibitemShut {NoStop}%
\bibitem [{\citenamefont {Essig}\ \emph {et~al.}(2019)\citenamefont {Essig},
  \citenamefont {P\'erez-R\'\i{}os}, \citenamefont {Ramani},\ and\
  \citenamefont {Slone}}]{Essig:2019kfe}%
  \BibitemOpen
  \bibfield  {author} {\bibinfo {author} {\bibfnamefont {Rouven}\ \bibnamefont
  {Essig}}, \bibinfo {author} {\bibfnamefont {Jes\'us}\ \bibnamefont
  {P\'erez-R\'\i{}os}}, \bibinfo {author} {\bibfnamefont {Harikrishnan}\
  \bibnamefont {Ramani}}, \ and\ \bibinfo {author} {\bibfnamefont {Oren}\
  \bibnamefont {Slone}},\ }\bibfield  {title} {\enquote {\bibinfo {title}
  {{Direct Detection of Spin-(In)dependent Nuclear Scattering of Sub-GeV Dark
  Matter Using Molecular Excitations}},}\ }\href {\doibase
  10.1103/PhysRevResearch.1.033105} {\bibfield  {journal} {\bibinfo  {journal}
  {Phys. Rev. Research.}\ }\textbf {\bibinfo {volume} {1}},\ \bibinfo {pages}
  {033105} (\bibinfo {year} {2019})},\ \Eprint
  {http://arxiv.org/abs/1907.07682} {arXiv:1907.07682 [hep-ph]} \BibitemShut
  {NoStop}%
\bibitem [{\citenamefont {Bunting}\ \emph {et~al.}(2017)\citenamefont
  {Bunting}, \citenamefont {Gratta}, \citenamefont {Melia},\ and\ \citenamefont
  {Rajendran}}]{Bunting:2017net}%
  \BibitemOpen
  \bibfield  {author} {\bibinfo {author} {\bibfnamefont {Philip~C.}\
  \bibnamefont {Bunting}}, \bibinfo {author} {\bibfnamefont {Giorgio}\
  \bibnamefont {Gratta}}, \bibinfo {author} {\bibfnamefont {Tom}\ \bibnamefont
  {Melia}}, \ and\ \bibinfo {author} {\bibfnamefont {Surjeet}\ \bibnamefont
  {Rajendran}},\ }\bibfield  {title} {\enquote {\bibinfo {title} {{Magnetic
  Bubble Chambers and Sub-GeV Dark Matter Direct Detection}},}\ }\href
  {\doibase 10.1103/PhysRevD.95.095001} {\bibfield  {journal} {\bibinfo
  {journal} {Phys. Rev. D}\ }\textbf {\bibinfo {volume} {95}},\ \bibinfo
  {pages} {095001} (\bibinfo {year} {2017})},\ \Eprint
  {http://arxiv.org/abs/1701.06566} {arXiv:1701.06566 [hep-ph]} \BibitemShut
  {NoStop}%
\bibitem [{\citenamefont {Chen}\ \emph {et~al.}(2020)\citenamefont {Chen},
  \citenamefont {Mahapatra}, \citenamefont {Agnolet}, \citenamefont {Nippe},
  \citenamefont {Lu}, \citenamefont {Bunting}, \citenamefont {Melia},
  \citenamefont {Rajendran}, \citenamefont {Gratta},\ and\ \citenamefont
  {Long}}]{Chen:2020jia}%
  \BibitemOpen
  \bibfield  {author} {\bibinfo {author} {\bibfnamefont {Hao}\ \bibnamefont
  {Chen}}, \bibinfo {author} {\bibfnamefont {Rupak}\ \bibnamefont {Mahapatra}},
  \bibinfo {author} {\bibfnamefont {Glenn}\ \bibnamefont {Agnolet}}, \bibinfo
  {author} {\bibfnamefont {Michael}\ \bibnamefont {Nippe}}, \bibinfo {author}
  {\bibfnamefont {Minjie}\ \bibnamefont {Lu}}, \bibinfo {author} {\bibfnamefont
  {Philip~C.}\ \bibnamefont {Bunting}}, \bibinfo {author} {\bibfnamefont {Tom}\
  \bibnamefont {Melia}}, \bibinfo {author} {\bibfnamefont {Surjeet}\
  \bibnamefont {Rajendran}}, \bibinfo {author} {\bibfnamefont {Giorgio}\
  \bibnamefont {Gratta}}, \ and\ \bibinfo {author} {\bibfnamefont {Jeffrey}\
  \bibnamefont {Long}},\ }\bibfield  {title} {\enquote {\bibinfo {title}
  {{Quantum Detection using Magnetic Avalanches in Single-Molecule Magnets}},}\
  }\href@noop {} {\  (\bibinfo {year} {2020})},\ \Eprint
  {http://arxiv.org/abs/2002.09409} {arXiv:2002.09409 [physics.ins-det]}
  \BibitemShut {NoStop}%
\bibitem [{\citenamefont {Berlin}\ \emph {et~al.}(2020)\citenamefont {Berlin},
  \citenamefont {D'Agnolo}, \citenamefont {Ellis}, \citenamefont {Schuster},\
  and\ \citenamefont {Toro}}]{Berlin:2019uco}%
  \BibitemOpen
  \bibfield  {author} {\bibinfo {author} {\bibfnamefont {Asher}\ \bibnamefont
  {Berlin}}, \bibinfo {author} {\bibfnamefont {Raffaele~Tito}\ \bibnamefont
  {D'Agnolo}}, \bibinfo {author} {\bibfnamefont {Sebastian A.~R.}\ \bibnamefont
  {Ellis}}, \bibinfo {author} {\bibfnamefont {Philip}\ \bibnamefont
  {Schuster}}, \ and\ \bibinfo {author} {\bibfnamefont {Natalia}\ \bibnamefont
  {Toro}},\ }\bibfield  {title} {\enquote {\bibinfo {title} {{Directly
  Deflecting Particle Dark Matter}},}\ }\href {\doibase
  10.1103/PhysRevLett.124.011801} {\bibfield  {journal} {\bibinfo  {journal}
  {Phys. Rev. Lett.}\ }\textbf {\bibinfo {volume} {124}},\ \bibinfo {pages}
  {011801} (\bibinfo {year} {2020})},\ \Eprint
  {http://arxiv.org/abs/1908.06982} {arXiv:1908.06982 [hep-ph]} \BibitemShut
  {NoStop}%
\bibitem [{\citenamefont {Pospelov}\ and\ \citenamefont
  {Ramani}(2020)}]{Pospelov:2020ktu}%
  \BibitemOpen
  \bibfield  {author} {\bibinfo {author} {\bibfnamefont {Maxim}\ \bibnamefont
  {Pospelov}}\ and\ \bibinfo {author} {\bibfnamefont {Harikrishnan}\
  \bibnamefont {Ramani}},\ }\bibfield  {title} {\enquote {\bibinfo {title}
  {{Earth-bound Milli-charge Relics}},}\ }\href@noop {} {\  (\bibinfo {year}
  {2020})},\ \Eprint {http://arxiv.org/abs/2012.03957} {arXiv:2012.03957
  [hep-ph]} \BibitemShut {NoStop}%
\end{thebibliography}%


%
\end{document}